\documentstyle[12pt,epsfig]{article}
\advance \topmargin by -\headheight
\advance \topmargin by -\headsep
     
\evensidemargin \oddsidemargin
\begin{document}
\topmargin -.6in
\def\br{\begin{eqnarray}}
\def\er{\end{eqnarray}}
\def\be{\begin{equation}}
\def\ee{\end{equation}}
\def\({\left(}
\def\){\right)}
\def\a{\alpha}
\def\b{\beta}
\def\d{\delta}
\def\D{\Delta}
\def\g{\gamma}
\def\G{\Gamma}
\def\h{ {1\over 2}  }
\def\hp{ {+{1\over 2}}  }
\def\hm{ {-{1\over 2}}  }
\def\k{\kappa}
\def\l{\lambda}
\def\L{\Lambda}
\def\m{\mu}
\def\n{\nu}
\def\o{\over}
\def\O{\Omega}
\def\p{\phi}
\def\rh{\rho}
\def\s{\sigma}
\def\t{\tau}
\def\th{\theta}
\def\ii {\'\i  }

\begin{center}
{\large {\bf Morse potential energy spectra through the variational\\ 
\vskip .5cm
method and supersymmetry}}\footnotemark
\footnotetext{PACS No. 31.15.Q, 11.30.P} 
\end{center}
\normalsize
\vskip 1cm
\begin{center}

{\it  Elso Drigo Filho $^a$ \footnotemark
\footnotetext{Work supported in part by CNPq} 
and Regina Maria  Ricotta $^b$} \\
$^a$ Instituto de Bioci\^encias, Letras e Ci\^encias Exatas, IBILCE-UNESP\\
Rua Cristov\~ao Colombo, 2265 -  15054-000 S\~ao Jos\'e do Rio Preto - SP\\
$^b$  Faculdade de Tecnologia de S\~ao Paulo, FATEC/SP-CEETPS-UNESP  \\
Pra\c ca  Fernando Prestes, 30 -  01124-060 S\~ao Paulo-SP\\ 
Brazil\\
\vskip 1cm
\end{center}
{\bf  Abstract}\\
The Variational Method is applied within the context of \-Supersymmetric Quantum Mechanics to
provide information about the energy and eigenfunction of the lowest levels of a Hamiltonian.
The approach is \-illustrated by the case of the Morse potential applied to several diatomic
molecules and the results are compared with stablished results.
\\

\noindent   {\bf I. Introduction}\\
Two decades ago Supersymmetric Quantum Mechanics, SQM, was born as a simplest case of field
theory in the study of SUSY breaking mechanism of higher dimensional quantum field theories,
developed to unify the four fundamental interactions in nature, namely the electroweak, strong
and gravitational interactions. So far, SQM has been extensively used to explore different
aspects of non-relativistic quantum mechanical systems (see, for instance,
\cite{Cooper1}-\cite{Haymaker} and references therein). The new algebraic method is  
specially  good to treat analytically solvable potential problems, \cite{Gedenshtein},
\cite{Drigo0}. Nevertheless, the approximation methods have shown to be of great interest to
provide new results in potential problems. In this context,  one has recently  suggested that the 
variational method could be applied within the formalism of SQM, \cite{Gozzi}-\cite{Drigo1}. This
approach has already been successfully applied to the Hulth\'en potential, \cite{Drigo1}. Here the
technique is applied to the Morse potential, one of the most successful models to represent the
states of diatomic molecules, \cite{ICooper}, \cite{Mazziotti} . The results are shown to be in
good agreamment with the numerical data and results from the other approximation techniques. 

The formalism of variational method can be found in several text books of quantum mechanics, 
(see for instance \cite{Schiff}); it can be used to obtain the approximate energy spectra of 
potentials,  in particular, the ground-state. The approach consists in using the trial
wave-function $\Psi(r)$ that depends of a number of parameters and in varying these parameters in the expression
for the expectation value of the energy 
\be
\label{energy}
E = {\int{\Psi^* H \Psi dr}\over  {\int{\mid \Psi \mid^2 dr}}}
\ee
where $H$ is the Hamiltonian of the system.  The variational parameters in $\Psi$ are varied until
the expectation value of the energy is minimum.  This value is an upper limit of the energy level. 
Usually this method is applied only to get the ground state energy, although it can be applied to
get the energy of the excited states as well.

Thus the central point of the variational method approach is the obtention of the trial
wavefunction.  At this crucial point the superalgebra is used to obtain this function. In
what follows, we briefly review the main principles of SQM and  introduce the variational
method. The case of angular momentum $l=0$, which is exactly solvable, is explicitly evaluated
from SQM formalism in order to get the hierarchy.  We then apply the variational scheme to the
Morse potential for different values of $l$ and for various diatomic molecules and compare the
results with known results from other numerical methods,
\cite{Bag}, \cite{Morales}.\\

\noindent {\bf  II. Supersymmetric Quantum Mechanics}\\

In SQM, \cite{Cooper1}-\cite{Drigo1}, for $N=2$ we have two
nilpotent operators, $Q$ and $Q^+$, that satisfy the algebra
\be
\{ Q, Q^+\} = H_{SS} \;\;\;\;; Q^2  = {Q^+}^2 = 0, 
\ee
where $H_{SS}$ is the supersymmetric Hamiltonian. This algebra can be realized as
\be
Q =  \left( \begin{array}{cc} 0  & 0  \\ A^-  & 0 
\end{array} \right ) \;,\;\;\;
Q^+ = \left( \begin{array}{cc} 0  & A^+  \\ 0 & 0 
\end{array} \right )
\ee
where $A^{\pm}$ are bosonic operators.  With this realization the supersymmetric Hamiltonian 
$H_{SS}$ is given by
\be
H_{SS} = \left( \begin{array}{cc} A^+A^-  &  0 \\ 0 & A^-A^+
\end{array} \right ) = \left( \begin{array}{cc} H^+  &  0 \\ 0 & H^-
\end{array} \right ).
\ee
$H^{\pm}$ are called supersymmetric partner Hamiltonians and  share the same spectra, apart
from the nondegenerate ground state.  Using the super-algebra a given Hamiltonian can be factorized
in terms of the bosonic operators. In $\hbar = c = 1$ units, it is given by
\be
H_1 =  -{1\o 2}{d^2 \o d r^2} + V_1(r) =  A_1^+A_1^-  + E_0^{(1)} 
\ee
where $ E_0^{(1)}$ is the lowest eigenvalue and the function $V_1(r)$ includes the barrier
potential term.  The bosonic operators are  defined by 
\be 
A_1^{\pm} =  {1\o \sqrt 2}\left(\mp {d \o dr} + W_1(r) \right) 
\ee
where the superpotential $W_1(r)$ satisfies the Riccati equation
\be
\label{Riccati}
W_1^2 - W_1'=  2V_1(r) - 2E_0^{(1)}  .    
\ee
The eigenfunction for the lowest state is related to the superpotential $W$ as
\be
\label{eigenfunction}
\Psi_0^{(1)} (r) = N exp( -\int_0^r W_1(\bar r) d\bar r).
\ee
Now it is possible to construct the supersymmetric partner Hamiltonian,
\be
H_2 = A_1^-A_1^+ + E_0^{(1)} =  -{1\o 2}{d^2 \o d r^2} + {1\o 2}(W_1^2 +
W_1')+ E_0^{(1)} .
\ee
If one factorizes  $H_2$ in terms of a new pair of bosonic operators,
$A_2^{\pm}$ one gets,
\be
H_2 = A_2^+A_2^- + E_0^{(2)} =  -{1\o 2}{d^2 \o d r^2} + {1\o 2}(W_2^2 -
W_2')+ E_0^{(2)} 
\ee
where $E_0^{(2)} $ is the lowest eigenvalue of $H_2$ and $W_2$ satisfy the
Riccati equation,
\be
W_2^2 - W_2'=  2V_2(r) - 2E_0^{(2)}  .
\ee
Thus a whole hierarchy of Hamiltonians can be constructed , with simple
relations connecting the eigenvalues and eigenfunctions of the $n$-members, 
\cite{Cooper1}, \cite{Sukumar1}-\cite{Drigo2}
\be
H_n = A_n^+A_n^- + E_0^{(n)} 
\ee
\be 
A_n^{\pm} =  {1\o \sqrt 2}\left(\mp {d \o dr} + W_n(r) \right) 
\ee
\be
\Psi_n^{(1)} = A_1^+A_2^+...\psi_0^{(n+1)}\;\;\;\;\;E_n^{(1)} = E_0^{(n+1)}
\ee
\be
\Psi_0^{(1)} (r) = N exp( -\int_0^r W_1(\bar r) d\bar r).
\ee

In this work our interest is in the ground-state eigenfunction,  to apply the
variational method. As shown above, in SQM formalism the ground state
eigenfunction can be determined from the superpotential  $W(r)$, which
satisfies the  Riccati equation, eq.(\ref {Riccati}).  Usually, if the
potential is not exactly solvable, it is easier to determine an
appro\-xi\-mation for the superpotential in eq.(\ref {Riccati}) than an
eigenfunction that satisfies the Schroedinger equation for the original
Hamiltonian. Thus we make an {\it ansatz} in the superpotential based  in the
superalgebra information in order to evaluate the trial wavefunction that
minimizes the expectation  value of the energy, eq.(\ref {energy}). This
formalism is applied to the Morse potential.\\

\vskip 1cm
\noindent {\bf III. The Morse Potential}\\

For the diatomic system, the three dimensional Morse oscillator can be written as,
\be
V_M = D (e^{-2a(r - r_e)} - 2e^{-a(r - r_e)})
\ee
where $D$ is the dissociation energy, $r_e$ is the equilibrium internuclear distance and $a$ is the
range parameter. We rewrite the original Schroedinger equation $H\Psi = E\Psi$ in terms of a new
varible
$y$,
\be
\left(-{d^2 \o d y^2} + {l(l+1) \o y^2} + \l^2( e^{-2(y - y_e)} - 2e^{-(y - y_e)})\right)\Psi(y) =
\epsilon \Psi(y)
\ee
where $y = ar$ and the constants are set like
\be
y_e = ar_e\;,\;\;\;\;\l^2 = {2m D \o a^2 \hbar^2}\;,\;\;\;\;E
= \epsilon {\hbar^2 a^2 \o 2m}
\ee 
and the parameter $m $ is the reduced mass of the molecule.\\

\noindent{\bf Case $l = 0$} \\ 
In this case the Schroedinger equation  is exactly solvable and therefore
the
$\,$ hierarchy of Hamiltonians can be constructed. For this case the Schroedinger equation is
reduced to the form
\be
\left(-{d^2 \o d y^2}  + \l^2( e^{-2(y - y_e)} - 2e^{-(y - y_e)})\right)\Psi(y) = \epsilon \Psi(y)
\ee
The associated Riccati equation is then
\be
\label{Riccati1}
W_1^2 - W_1' + \epsilon_0^{(1)} = V_1(y) 
\ee
with 
\be
V_1(y) = \l^2( e^{-2(y - y_e)} - 2e^{-(y - y_e)})
\ee
whose solution is given by 
\be
\label{W_1}
W_1(y) = - \lambda e^{-(y - y_e)} + (\lambda - 1/2)
\ee
and the ground-state energy is 
\be
\epsilon_0^{(1)} = -(\lambda - 1/2)^2
\ee
To find the second member of the super-family we solve the equation
\be
W_1^2 + W_1' + \epsilon_0^{(1)} = V_2(r)
\ee
We then find that the superpartner is given by
\be
W_2(y) = - \lambda e^{-(y - y_e)} + (\lambda - 3/2)
\ee
and the ground-state energy is 
\be
\epsilon_0^{(2)} = - (\lambda - 3/2)^2.
\ee
The process of factorization can be continued and the whole hierarchy can be  evaluated. The
result is
\br
V_{n+1}(y) = \l^2( e^{-2(y - y_e)} - 2e^{-(y - y_e)}) + 2n \lambda e^{-(y - y_e)}
\nonumber
\er
\be
W_{n+1}(y) = - \lambda e^{-(y - y_e)} + (\lambda - {2n + 1\o 2})  
\ee
\br
\epsilon_0^{(n+1)} = -(\lambda - {2n + 1\o 2})^2. \nonumber
\er
\\
\noindent {\bf Case $l\not=0$} \\
In this case an analytical exact soluction cannot be determined. 
Nevertheless, based in the above arguments on how to obtain the trial wavefunction, we make the
following {\it ansatz}  for the superpotential
\be
\label{W_1(C)}
W_1(y) = -\l e^{-(y - y_e)} - {(l+1) \o y} + C.
\ee
The first term is taken from the one-dimensional Morse superpotential, case of $l=0$, 
eq.(\ref{W_1}),  \cite{Drigo3}.  The knowledge of the second term comes from the study of
three-dimensional potentials, (see, for instance, assyntotically linear potential, \cite{Drigo4},
and the truncated Coulomb potential,
\cite{Drigo5}).  The $c$-number $C$ will be taken as the variational parameter in the trial
wavefunction.

The eigenfunction obtained from eq. (\ref{eigenfunction}) is then
\be
\Psi(y) \propto e^{-\l e^{-(y - y_e)}}\;  y^{l+1} \;  e^{-Cy} 
\ee

Using this expression as a trial wavefunction in the variational method we change the parameter 
$C$ by the variational parameter $\mu$, i.e.,
\be
\Psi_{\mu} = \Psi(y, C = \mu) \propto e^{-\l e^{-(y - y_e)}}\;  y^{l+1} \;
e^{-\mu y}.
\ee
The energy  is then obtained by minimisation of the energy expectation value  with respect to $\mu$.
Thus, the equation to be  minimised is  
\be
\label{energymu}
E_{\mu} = {\int_0^{\infty} \Psi_{\mu}(y) [- {d^2 \o dy^2} + \l^2( e^{-2(y - y_e)} - 2e^{-(y -
y_e)}) + {l(l+1)\o y^2}] \Psi_{\mu}(y) dy
\o \int_0^{\infty} \Psi_{\mu}(y)^2 dy}.
\ee

We have used this expression to minimize the energy expectation value of
various molecules: $H_2$, $HCl$, $CO$ and $LiH$.  The explicit values of the
energy  for $n=0$ and different values of $l$ are shown, for 
known values of their respective potential  parameters, \cite{Varshni}: $D$, $a$, $r_e$ and
$m$. \\
\vskip .5cm
{\bf Table 1.} Energy eigenvalues (in $eV$) for different values of $l$ for
$H_2$ molecule, with $D = 4.7446 eV$, $a = 1.9426 \AA^{-1}$, $r_e = 0,7416
\AA$ and
$m = 0,50391amu$. Comparison is made with results from  ref.\cite{Morales}.\\
\vskip .5cm
\begin{tabular}{|l|c|c|c|c|c|} \hline
\multicolumn{1}{|c} {l} &
\multicolumn{1}{|c} {Variational results }   &
\multicolumn{1}{|c|} {Shifted $1/N$ } & 
\multicolumn{1}{|c|} {Modified shifted $1/N$ }  &
\multicolumn{1}{|c|} {Exact Numerical} \\  
\multicolumn{1}{|c} {} &
\multicolumn{1}{|c} {} & 
\multicolumn{1}{|c} {expansion results } &
\multicolumn{1}{|c} {expansion results } &
\multicolumn{1}{|c|} {}  \\ \hline
 0 & -4.4758  & -4.4749  & -4.4760 & -4.4759 \\ \hline  
 5 & -4.2563  & -4.2589  & -4.2592 & -4.2589\\ \hline 
10 & -3.7187  & -3.7247  & -3.7252 & -3.7242\\ \hline 
 15 & -2.9578  & -2.9663 & -2.9670 & -2.9654\\ \hline 
20 & -2.0735   & -2.0839 & -2.0846 & -2.0826\\ \hline 
\end{tabular}\\
\vskip 1cm
{\bf Table 2.} Energy eigenvalues (in $eV$) for different values of $l$ for
$HCl$ molecule, with $D = 37255 cm^{-1}$, $a = 1.8677 \AA^{-1}$, $r_e =
1.2746 \AA$ and
$m = 0.9801045 amu$. Comparison is made with results from  ref.\cite{Bag}.  \\
\vskip .5cm
\begin{tabular}{|l|c|c|c|c|c|} \hline
\multicolumn{1}{|c} {l} &
\multicolumn{1}{|c} {Variational results }   &
\multicolumn{1}{|c|} {Shifted $1/N$ } & 
\multicolumn{1}{|c|} {Modified shifted $1/N$ }  \\  
\multicolumn{1}{|c} {} &
\multicolumn{1}{|c} {} &
\multicolumn{1}{|c} {expansion results } &
\multicolumn{1}{|c|} {expansion results } \\ \hline
 0 & -4.4360  & -4.4352 & -4.4355   \\ \hline  
 5 & -4.3971  & -4.3967 & -4.3968   \\ \hline 
10 & -4.2940  & -4.2939 & -4.2940  \\ \hline 
\end{tabular}\\
\vskip 1cm
{\bf Table 3.} Energy eigenvalues (in $eV$) for different values of $l$ for
$CO$ molecule, with $D = 90540 cm^{-1}$, $a = 2.2994 \AA^{-1}$, $r_e = 1.1283
\AA$ and
$m = 6.8606719 amu$. Comparison is made with results from  ref.\cite{Bag}.  \\
\vskip .5cm
\begin{tabular}{|l|c|c|c|c|c|} \hline
\multicolumn{1}{|c} {l} &
\multicolumn{1}{|c} {Variational results}   &
\multicolumn{1}{|c|} {Shifted $1/N$ } & 
\multicolumn{1}{|c|} {Modified shifted $1/N$ }  \\  
\multicolumn{1}{|c} {} &
\multicolumn{1}{|c} {} &
\multicolumn{1}{|c} {expansion results} &
\multicolumn{1}{|c|} {expansion results} \\ \hline
 0 & -11.093  & -11.091 & -11.092   \\ \hline  
 5 & -11.085  & -11.084 &-11.084   \\ \hline 
10 & -11.066  & -11.065 & -11.065  \\ \hline 
\end{tabular}\\
\vskip 1cm
{\bf Table 4.} Energy eigenvalues (in $eV$) for different values of $l$ for
$LiH$ molecule, with $D =20287cm^{-1}$, $a = 1.1280 \AA^{-1}$, $r_e = 1.5956
\AA$ and
$m = 0.8801221 amu$. Comparison is made with results from  ref.\cite{Bag}. \\
\vskip .5cm 
\begin{tabular}{|l|c|c|c|c|c|} \hline
\multicolumn{1}{|c} {l} &
\multicolumn{1}{|c} {Variational results }   &
\multicolumn{1}{|c|} {Shifted $1/N$ } & 
\multicolumn{1}{|c|} {Modified shifted $1/N$ }  \\  
\multicolumn{1}{|c} {} &
\multicolumn{1}{|c} {} &
\multicolumn{1}{|c} {expansion results } &
\multicolumn{1}{|c|} {expansion results } \\ \hline
 0 & -2.4291  & -2.4278 & -2.4280   \\ \hline  
 5 & -2.4014  & -2.3999 & -2.4000   \\ \hline 
10 & -2.3287  & -2.3261 & -2.3261  \\ \hline 
\end{tabular}\\
\vskip 1cm 
We stress that in fact we are dealing with an effective potential $V_{eff}$ that satisfies
Riccati equation, i.e., 
\be
V_{eff}(y) = \bar W_1^2 - \bar W_1'+ E(\bar\mu)    
\ee
where $ \bar W_1 = W_1(\bar\mu)$ is given by eq.(\ref{W_1(C)}) and  $\bar\mu$ is the parameter
that minimises the energy of eq.(\ref{energymu}). It is given by
\be
V_{eff} = -\l e^{-(y - y_e)} + \left( -\l e^{-(y - y_e)} + \mu - {l+1 \o y}\right )^2 -  {l+1 \o
y} + E(\bar\mu)
\ee 
The  plots of both $V_{eff}$ and $V_M$ (plus the potential barrier term) in the same graph for
$l=5$ are shown bellow. The upper curve corresponds to $V_{eff}$. We notice that for low energies 
the effective potential is quite similar
to the real Morse potential and this is why our results are in such a good agreement with the
other results. 

\begin{figure}[h]
\centering
\includegraphics[width=8cm]{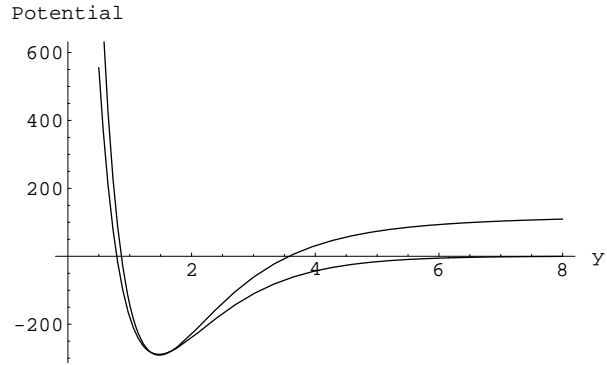}
\caption{Plot of $V_M$ plus the potential barrier term and $V_{eff}$.}
\label{Figura1}
\end{figure}

\vskip 1cm
{\bf IV. Conclusions}\\

The SQM formalism was used to explore the Morse potential. The hierarchy of Hamiltonians was
evaluated for the case of angular momentum $l = 0$, for which the Morse potential is exactly
solvable.

We have  shown a new approach through the application of the variational
method with the SQM formalism to evaluate the eigenvalues of the Morse
oscillator for $\;\;$ different values of the angular momentum $l$.  The determination of the
trial wave function motivated by SQM was made by an {\it ansatz} in the superpotential.  The 
application of the method for diatomic molecules indicated that the values obtained are in very
good agreement with shifted large-$N$ tecnique results, (SLNT), modified large-$N$ tecnique
results (SLNT) and numerical data. 

We conclude that the conception of the trial wavefunction through the superpotential is a simple
and good alternative procedure that enables to evaluate the energy eigenvalues with a reasoble
accuracy. This is  not a priori obvious and it suggests that it is a suitable method to treat
other potentials. 

{\bf Acknowledgements}\\
The authors would like to thank Prof. U. P. Sukhatme for reading the manuscript.
 

\begin{thebibliography}{99}                                            
\bibitem{Cooper1} F. Cooper, A. Khare and U. P. Sukhatme, Phys. Rep. {\bf 251} (1995) 267
\bibitem{Haymaker} R. W. Haymaker and A. R. Rau,  Am. J. Phys. {\bf 54} (1986) 928
\bibitem{Gedenshtein} L. Gedenshtein and I. V. Krive, So. Phys. Usp. {\bf 28} (1985) 645
\bibitem{Drigo0} E. Drigo Filho and R. M. Ricotta, hep-th 9904038
\bibitem{Gozzi} E. Gozzi, M. Reuter and W. D. Thacker, Phys. Lett. {\bf A183} (1993) 29
\bibitem{Cooper2} F. Cooper, J. Dawson and H. Shepard, Phys. Lett. {\bf 187A} (1994) 140
\bibitem{Drigo1} E. Drigo Filho and R. M.
Ricotta, Mod. Phys. Lett. {\bf A10} (1995) 1613 
\bibitem{ICooper} I. L. Cooper, J. Phys. Chem. {\bf A 102} (1998) 9565
\bibitem{Mazziotti} D. A. Mazziotti, Chem. Phys. Lett. {\bf 299} (1999) 473
\bibitem{Schiff} L. I. Schiff, {\it Quantum Mechanics}, McGraw-Hill Book Co.,
NY, 1968
\bibitem{Bag} M. Bag, M. M. Panja, R. Dutt  and Y. P. Varshni, Phys. Rev.
{\bf A46} (1992) 6059 
\bibitem{Morales} D. A. Morales, Chem. Phys. Lett. {\bf 161} (1989) 253
\bibitem{Sukumar1} C. V. Sukumar, J. Phys. A: Math. Gen. {\bf 18} (1985) L57
\bibitem{Sukumar2} C. V. Sukumar, J. Phys. A: Math. Gen. {\bf 18} (1985) 2917
\bibitem{Drigo2} E. Drigo Filho and R. M. Ricotta, Mod. Phys. Lett. {\bf A14} (1989) 2283
\bibitem{Drigo3} E. Drigo Filho, J. Phys. A: Math. Gen. {\bf 21} (1988) L1025
\bibitem{Drigo4} E. Drigo Filho, Mod. Phys. Lett. {\bf A11} (1996) 207
\bibitem{Drigo5} E. Drigo Filho, Mod. Phys. Lett. {\bf A9} (1994) 411
\bibitem{Varshni} Y. P. Varshni, Can. J. Chem. {\bf 66} (1988) 763
\end{thebibliography}
\end{document}